\documentclass[10pt]{article}
\usepackage[dvips]{graphicx}

\usepackage{epsfig}
\usepackage{graphicx}
\usepackage{indentfirst}
\usepackage{amsmath}
\usepackage{amsfonts}
\usepackage{amssymb}

\begin{document}
\begin{center}
{\Large\bf   Cylindrical Solutions in Modified f(T) Gravity.\\}
\medskip

M. J. S. Houndjo$^{a,b}$\footnote{e-mail:
sthoundjo@yahoo.fr},\, D. Momeni\footnote{e-mail: d.momeni@yahoo.com}$^{c}$\,and R. Myrzakulov$^{c}$\footnote{e-mail:rmyrzakulov@csufresno.edu}\\
$^{a}${  Departamento de Engenharia e Ci\^{e}ncias Naturais- CEUNES -
Universidade Federal do Esp\'irito Santo\\
CEP 29933-415 - S\~ao Mateus - ES, Brazil}\\
 $^{b}${ Institut de Math\'{e}matiques et de Sciences Physiques (IMSP) - 01 BP 613 Porto-Novo, B\'{e}nin}\\
$^{c}${
Eurasian International Center for Theoretical Physics - 
 Eurasian National University,\\ Astana 010008, Kazakhstan}\\

\date{}

\end{center}
\begin{abstract}
We investigate static cylindrically symmetric vacuum solutions in Weyl coordinates in the framework of $f(T)$ theory of gravity, where $T$ is the torsion scalar. The set of modified Einstein equations is presented  and the fourth coming equations are established. Specific physical expressions are assumed for the algebraic function $f(T)$ and solutions are obtained. Moreover, general solution is obtained with finite values of $u(r)$ on the axis $r=0$, and this leads to a constant torsion scalar. Cosmological constant is also introduced and its relation to Linet-Tian solution in GR is commented.
\end{abstract}

Pacs numbers: 04.50.kd, 04.20.-q
\section{Introduction}
The accelerated expansion of the present universe is widely  accepted by many observational evidence \cite{moha1}-\cite{moha15}. Many schemes are proposed to explain this phenomenon. The dark energy scenario \cite{rui1}-\cite{rui13} is the most popular one among them. Also, the models based on infra-red modifications to General Relativity (GR), such as scalar tensor theories \cite{rui2,rui22} have been considered, as well as $f(R)$ gravity \cite{rui3,rui31} gravity and braneworld models \cite{rui4}. The main problem that remains is that the resulting field equations are fourth order because the Ricci scalar is constructed from the second order derivatives of the metric, and  this feature leads to pathologies.\par  
Recently, an alternative to GR is the so-called $f(T)$  theory, which has received considerable attention as a possible explanation of the late-time acceleration of the universe \cite{james30}-\cite{james32}. This theory is the modification of the `` Teleparallel" equivalence of GR, (TEGR) for which, instead of using Riemann Cartan space with curvature but no torsion, uses the Weitzenb$\ddot{\mbox{o}}$ck connection that has no curvature but only torsion. \par
Static solutions in $f(T)$ with spherical symmetry have been investigated and interesting results have been found \cite{daouda1,daouda2}. Note that cylindrical symmetry is the next symmetry considered normally in the study of exact solutions in GR (not just for theoretical reasons but also because they might have physical realization in objects such as cosmic strings). Hence, it seems natural to extend the studies of exact solutions in $f(T)$ theory in the same way. In this paper, we propose to find static solutions in $f(T)$ theory, but with cylindrical symmetry.\par
The outline of the paper is as follows. In Sec. $2$ we present the formalism of the theory and the general field equations. The tetrads basis for cylindrical metric is explored and the four equations are established from the general one, in Sec. $3$. In Sec. $4$, the vacuum solution are obtained from some assumptions of physical algebraic function $f(T)$. The discussions and conclusion are presented in the last section, Sec. $5$.

\section{Formalism of $f(T)$ gravity}

Gauge theory of gravity is based on the  equivalence principle.
For example, $SL(2,C)$ gauge theory on the gravitational field can
be used for quantization of this fundamental force \cite{carmeli2}.
We are working with a curved manifold for the construction of a
gauge theory for gravitational field. It is not necessary to use
only the Riemannian manifolds. The general form of a gauge theory
for gravity, with metric, non-metricity and torsion can be
constructed easily \cite{smalley}. If we relax the non-metricity,
our theory is defined on Weitzenb\"{o}ck spacetime, with torsion and
with zero local Riemann tensor $R_{\alpha\beta\gamma}^{\delta}=0$.
In this theory, which is called teleparallel gravity, we use a
non-Riemannian spacetime manifold. The dynamics of the metric
determined using the torsion scalar $T$. The basic quantities in
teleparallel theory or its  natural extension, namely $f(T)$ gravity,
are the vierbein (tetrad) basis  $e^{i}_{\mu}$
\cite{ff}-\cite{darabi10}. This basis is an orthonormal coordinate
free basis, defined by the following equation
\begin{eqnarray}\nonumber
g_{\mu\nu}=e_{\mu}^{i}e_{\nu}^j \eta_{ij}\,,
\end{eqnarray}
where $\eta_{ij}$ is the
Minkowski flat tensor. Since the basis is orthonormal, one gets $e^{i}_{\mu}e^{\mu
}_j=\delta^i_{j}$.
A suitable form of the action for $f(T)$ gravity in Weitzenb\"{o}ck
spacetime is given by \cite{darabi}-\cite{darabi10}
\begin {equation}\label{a-1}
S=\int d^{4}xe\Big(\frac{1}{16\pi}f(T)+L_{m}\Big)\,,
\end{equation}
where  $f$ is an arbitrary function of $T$, and $e=\det(e^{i}_{\mu})$. Here $T$ is defined by
\begin{equation}\nonumber
T=S^{\:\:\:\mu \nu}_{\rho} T_{\:\:\:\mu \nu}^{\rho}\,,
\end{equation}
with
$$
T_{\:\:\:\mu \nu}^{\rho}=e_i^{\rho}(\partial_{\mu}
e^i_{\nu}-\partial_{\nu} e^i_{\mu})\,,
$$
$$
S^{\:\:\:\mu \nu}_{\rho}=\frac{1}{2}(K^{\mu
\nu}_{\:\:\:\:\:\rho}+\delta^{\mu}_{\rho} T^{\theta
\nu}_{\:\:\:\theta}-\delta^{\nu}_{\rho} T^{\theta
\mu}_{\:\:\:\theta})\,,
$$
where the asymmetric tensor (the so-called contorsion tensor) $K^{\mu \nu}_{\:\:\:\:\:\rho}$ reads
$$
K^{\mu \nu}_{\:\:\:\:\:\rho}=-\frac{1}{2}(T^{\mu
\nu}_{\:\:\:\:\:\rho}-T^{\nu \mu}_{\:\:\:\:\:\rho}-T^{\:\:\:\mu
\nu}_{\rho})\,.
$$
The equation of motion derived from the action, by varying the action with respect to  $e^{i}_{\mu}$,
is given by
\begin{eqnarray}
S^{\;\;\nu\rho}_{\mu}\partial_{\rho}Tf_{TT}+\left[e^{-1}e^{i}_{\mu}\partial_{\rho}\left(ee^{\;\;\alpha}_{i}S^{\;\;\nu\rho}_{\alpha}\right)+T^{\alpha}_{\;\;\lambda\mu}S^{\;\;\nu\lambda}_{\alpha}\right]f_{T}+\frac{1}{4}\delta^{\nu}_{\mu}f=4\pi\mathcal{T}^{\nu}_{\mu}\,\,\,.\label{eom}
\end{eqnarray}
where
$\mathcal{T}_{\mu\nu}=e^{a}_{\mu}\mathcal{T}_{a\nu}$ is the energy-momentum tensor for matter sector of the Lagrangian $ L_m$, and is defined by
$$
\mathcal{T}_{a\nu}=\frac{1}{e}\frac{\delta L_m}{\delta e^{a\nu}}\,\,.
$$
The covariant derivatives are compatible with the metricity
$g^{\mu\nu}_{\,\,\,;\mu}=0$. It is a straightforward calculation to show
that  (\ref{eom}) is reduced to the Einstein gravity when $f(T)=T$.
This is the equivalence between the teleparallel theory and the
Einstein gravity \cite{T}-\cite{T2}. Note that teleparallel gravity is not
unique, since it can either be described by any Lagrangian which
remains invariance under the local or global Lorentz $SO(3, 1)$
group \cite{tegr-tegr1}.\par

We mention here that a general Poincare gauge invariance model for
gravity (the so-called Einstein-Cartan-Sciama-Kibble (ECSK)) is
previously reported in the literatures \cite{ECSK}. Specially, in
the framework of the Poincar\'e gauge invariant form of the
ECSK theory, the \textit{notions of ``dark matter'' and ``dark
energy'' play a role similar to that of ``ether'' in physics
before the creation of special relativity theory}
\cite{plb}-\cite{plb2}.  In this Letter, we focus only on $f(T)$ models, without
curvature and with non zero torsion.

\section{Tetrads basis for cylindrical metric}
The metric of a cylindrically symmetric space-time in the Weyl static gauge reads \cite{stephani}
\begin{eqnarray}\label{g}
g_{\mu\nu}=diag\Big[e^{2u},\,-e^{2(k-u)},\,-w^2e^{-2u},\,-e^{2(k-u)}\Big],\ \ x^{\mu}=(t,r,\varphi,z)\,\,.
\end{eqnarray}
All of the metric is functions of the radial coordinate $r$. One simple tetrad basis is
\begin{eqnarray}
e^{i}_{\mu}=diag\Big[ e^u,\, e^{k-u},\, we^{-u},\, e^{k-u} \big]\,\,.\\
e=det(e^{i}_{\mu})=w e^{2(k-u)}\,\,.
\end{eqnarray}
The non null components of the torsion tensor $T^{\rho}_{\,\,\,\mu\nu}$ read
\begin{eqnarray}
T^{0}_{\,\,\,\mu\nu}=-u'\delta_{\mu 0}\delta_{\nu 1}\,\,,\\
T^{2}_{\,\,\,\mu\nu}=   \frac{wu'-w'}{w}\delta_{\mu 2}\delta_{\nu 1}\,\,,\\
T^{3}_{\,\,\,\mu\nu}=(k'-u')\delta_{\mu 1}\delta_{\nu 3}\,\,\,.
\end{eqnarray}
The non null components of the contorsion $K^{\mu \nu}_{\:\:\:\:\:\rho}$ are \begin{eqnarray}
 K^{01}_{\:\:\:\:\:0}=u'e^{2(u-k)}\,\,,\\
 K^{12}_{\:\:\:\:\:2}=\frac{e^{2(u-k)}}{w}(wu'-w')\,\,,\\
 K^{13}_{\:\:\:\:\:3}=e^{2(u-k)}(u'-k')\,\,\,,
\end{eqnarray}
and the non null components of the tensor $S$ read

\begin{eqnarray}
S^{\:\:\:01}_{0}=\frac{e^{2(u-k)}}{2w}\Big[2wu'-w'-wk'\Big]\,\,,\\
S^{\:\:\:12}_{2}=\frac{k'}{2}e^{2(u-k)}\,\,,\\
S^{\:\:\:13}_{3}=\frac{w'}{2w}e^{2(u-k)}\,\,.
\end{eqnarray}
The torsion scalar reads
\begin{eqnarray}
T=\frac{e^{2(u-k)}}{w}\Big[2wu'^2-2k'w'\Big]\,\,.\label{t}
\end{eqnarray}
 Now the field equations by assuming a diagonal perfect fluid with energy momentum tensor
\begin{eqnarray}
 T_{\mu\nu}=diag(\rho,-p_r,-p_\varphi,-p_z)\,\,\,
\end{eqnarray}
 read
 \begin{eqnarray}\label{eq1}
4\pi\rho=\frac{f}{4}+\frac{e^{2(u-k)}}{2w}\Big(2wu''+2w'u'-w''-k'w'-wk''\Big)f_T-\frac{e^{4(u-k)}}{w^3}f_{TT}\times\\ \nonumber
\Big[4w^3u'^2u''-2w^2w'u'u''-2w^3k'u'u''+4w^3u'^4-2w'w^2u'^3-6w^3k'u'^3+2w^3k'^2u'^2-2k'w'w^2u'^2\\ \nonumber -2k'w''w^2u'+4wk'u'w'^2+6w'u'k'^2w^2-2w'w^2k''u'+wk'w'w''+w^2k'^2w''\\ \nonumber
-k'w'^3+ww'^2k''-3wk'^2w'^2+w^2w'k'k''-2w^2w'k'^3
\Big]\,\,,
\end{eqnarray}

\begin{eqnarray}
-4\pi p_r=\frac{f}{4}+\frac{e^{2(u-k)}}{w}(wu'^2-k'w')f_{T}\,\,,\label{eq2}\\
-4\pi p_{\varphi}=\frac{f}{4}+\frac{e^{2(u-k)}}{2w}(k'w'+wk'')f_{T}+\frac{e^{4(u-k)}}{w^2}f_{TT}\times\\ \nonumber \Big[2k'w^2u'u''+2k'w^2u'^3-2k'^2w^2u'^2-2k'^2ww'u'-k'^2ww''+w'^2k'^2+2k'^3ww'-wk'w'k''\Big]\,\,,\label{eq3}
\end{eqnarray}

\begin{eqnarray}\label{eq4}
-4\pi p_z=\frac{f}{4}-\frac{w''}{2w}e^{2(u-k)}f_{T}+\frac{e^{4(u-k)}}{w^3}f_{TT}\times\\ \nonumber \Big[2w^2w'u'u''+2w^2w'u'^3-2k'w^2w'u'^2-2wk'w'^2u'-wk'w'w''+k'w'^3+2k'^2ww'^2-ww'^2k''\Big]\,\,.
\end{eqnarray}

\section{Vacuum solutions}
In this section we analyze the vacuum solution , i.e $\rho=p_r=p_{\varphi}=p_z=0$.
First, note that the system (\ref{eq1}-\ref{eq4}) in the case of the TEGR , $f(T)=T$, reduces to the field equations in the literature \cite{stephani}. In vacuum, we are searching for solutions for metric functions $\{u,k,w\}$ and known forms of the $f(T)$. First we discuss the TEGR case.

\subsection{Solution with $f(T)=T$}
In this case we obtain

\begin{eqnarray}
k(r)=c_5-\frac{1}{c_3}\log(r+c_4),\ \ c_3<0\,\,,\\
u(r)=\pm \frac{\log(r+c_4)}{\sqrt{-c_3}}\,\,,\\
w(r)=-c_1c_3(r+c_4)\,\,.
\end{eqnarray}
It could be seen that this is a Torsionless solution (i.e. $T = 0$ in (\ref{t})) in
which we should identify which one of the constants $c_1$ to $c_5$ corresponds
to physical parameters of the spacetime and which ones
could be absorbed into the coordinate redefinitions. Substituting
the above functions back into the metric form (\ref{g}), we obtain
\begin{eqnarray}
ds^2=\Big[\frac{\rho}{-c_1c_3}\Big]^{\pm\frac{2}{\sqrt{-c_3}}}dt^2-e^{2c_5}\Big[\frac{\rho}{-c_1c_3}\Big]^{\tilde{c_3}}(\frac{d\rho^2}{c_1^2c_3^2}+dz^2)-\rho^{2(1\mp\frac{1}{\sqrt{-c_3}})}(-\frac{1}{c_1c_3})^{\mp \frac{2}{\sqrt{-c_3}}}d\varphi^2\,\,, \label{line1}
\end{eqnarray}

in which $\tilde{c_3}=\frac{2}{-c_3}\mp \frac{2}{\sqrt{-c_3}}$
 and $\rho=w(r)=-c_1c_3r$ is the new radial coordinate
by setting $c_4 = 0$ without any loss of generality.  First, we must decide about the mass parameter of the solution. In fact, it means {\it mass per length} of the cylindrical spacetime. It will be clear by performing a coordinate transformation on the radius coordinate $\rho$ to a new one $\tilde{\rho}$. Consider the following correspondence between the two pair of the coordinates $(\rho,\tilde{\rho})$:
\begin{eqnarray}
e^{2c_5}\Big[\frac{\rho}{-c_1c_3}\Big]^{\tilde{c_3}}\frac{d\rho^2}{(-c_1c_3)^2}=\tilde{\rho}^{2m(m\pm1)}d\tilde{\rho}^2
\end{eqnarray}
We identify $m(m\pm1)=\frac{\tilde{c_3}}{2}=\frac{1}{-c_3}\mp \frac{1}{\sqrt{-c_3}}$ to recovering the common form of the vacuum solution of the cylindrical symmetry. The unique solution for $m$ reads
\begin{equation}
m=\frac{1}{\sqrt{-c_3}}.
\end{equation}
This is the parametrized mass per length of the vacuum solution. With this value of $m$ we can obtain the relation between two radial coordinates
\begin{eqnarray}
\tilde{\rho}=A^{\frac{1}{1+m(m\mp1)}},\ \ A=\frac{e^{c_5}}{(-c_1c_3)^{1+m(m\mp1)}}.
\end{eqnarray}
Now, by replacing the ${tt}$ component of the metric in terms of the $\tilde{\rho}$ we observe that it is adequate to redefine the time in the following form
\begin{eqnarray}
\tilde{t}=\frac{A^{\pm\frac{m}{1+m(m\mp1)}}}{(-c_1c_3)^{\pm m}}t.
\end{eqnarray}
Finally, for the spatial coordinate $z$ and the azimuthal coordinate $\varphi$ we obtain
\begin{eqnarray}
\tilde{z}=e^{c_5}\frac{A^{-\frac{m(m\mp1)}{1+m(m\mp1)}}}{(-c_1c_3)^{m(m\mp1)}}z,
\\
\tilde{\varphi}=A^{\frac{1\mp m/2}{1+m(m\mp1)}}\varphi.
\end{eqnarray}

and the above metric reduces to
\begin{eqnarray}
ds^2=\tilde{\rho}^{\pm 2m}d\tilde{t}^2-\tilde{\rho}^{2\mp m}d\tilde{\varphi}^2-\tilde{\rho}^{2m(m\mp 1)}(d\tilde{\rho}^2+d\tilde{z}^2)\,\,,
\end{eqnarray}
which is formally similar to the Levi-Civita \cite{LC} static cylindrically
symmetric solution in GR normally written without $\pm$ sign but
with the constant $m$ taking both positive and negative values. It
should also be noted that the range of the variable $m$ is not in
general $(0, 2\pi]$, not even for the flat cases of $m = 0, 1$.
In the case of $m = 0$ the spacetime is conical with a deficit
angle corresponding to the exterior metric of a cosmic string with
the following line element \cite{vilenkin}
\begin{eqnarray}
ds^2=d\tilde{t}^2-a_0^2\tilde{\rho}^{2}d\tilde{\varphi}^2-(d\tilde{\rho}^2+d\tilde{z}^2)\,\,,
\end{eqnarray}
in which $a_0 = -c_1c_3 e^{-c_5}$ is the conical parameter related to the gravitational
mass per unit length of the spacetime, as 

\begin{eqnarray}
a_0 =1-4\eta\,\,,
\end{eqnarray}

such that $0 < a_0 < 1$ for $0 < c_5 <\infty$ (taking $c_1c_3 = 1$). This metric,
exposing the geometry around a straight cosmic string, is locally
identical to that of flat spacetime, but is not globally Euclidean
since the angle $\tilde{\varphi}=a_0\varphi$ varies in the range $0\leq\tilde{\varphi}<2\pi(1-\eta) $. The same solution reported previously in $f(R)$ gravity \cite{plb}-\cite{plb2}.

\subsection{Solutions for $f(T)=T+\beta T^2$}
 In this section we would like to derive the general exact solution for a viable $f(T)$ model, with a quadratic correction term, i.e. $f(T)=T+\beta T^2$. The motivation for considering this model for the gravity sector of our model will be discussed here.\\
Let us consider a generic form of the $f(T)$ model, without any singularity in $T=0$. It is clear that we can expand the $f(T)$ function as a Taylor series up to order of the $O(T^2)$ in the following form
\begin{eqnarray}
f(T)=f(0)+Tf'(0)+\frac{T^2F''(0)}{2}
\end{eqnarray}
It is obvious that the first term $f(0)$ can be adjusted by the cosmological constant term  $\Lambda$. But in this section we want just to obtain the vacuum solutions without  $\Lambda$. Therefore, we just put $\Lambda\sim f(0)=0$. The next non zero term can be interpreted as the TEGR term, so we set $f'(0)=1$ in units of the $2\kappa^2=1$. The third term is of second order, {\it leading order} term, which can be written in the form of $\beta=\frac{f''(0)}{2}$. Thus, we take the $f(T)$ model as the following term
\begin{eqnarray}
f(T)=T+\beta T^2
\end{eqnarray}
 In the above expression, $\beta$ is a pure geometrical
parameter that determines at first approximation the divergence
from teleparallel gravity, that is from General Relativity. As we know, in the Schwarzschild metric in the weak field approximation, $\beta\sim \Lambda$. Also,
the above simple ansatz has been used as a first non-linear correction of
$f(T)$ gravity . \par
In particular, considering the basic and
usual ansatz $f(T)=T+\beta T^2$ is a good approximation in all realistic
cases and we are able to use data from planetary
motions in order to constrain $\beta$ \cite{saridakis}. Further, we mention here that in the limit $\beta=0$ the above
model coincides with the TEGR. It is important to recall that one can also verify that the quadratic correction to the TEGR is very small, i.e. in fact $\beta T^2<<T$. So, we can apply perturbations scheme if it is needed.

In this case field equations have the following exact solutions
\begin{eqnarray}
u(r) =c_1,\ \ k(r)=\frac{c_1^2}{2}r^2+c_2+\frac{1}{2}\log(\frac{5\beta}{-c_2-\frac{3r^2}{2}e^{-2c_1}}),\ \ w(r)=c_0 r \,\,.
\end{eqnarray}
This is an exact solution which has non constant torsion, indeed from (\ref{t}) we obtain
$$
T=-\,{\frac {3{{\rm e}^{-2\,c_{{2}}-{c_{{1}}}^{2}{r}^{2}}}}{5\beta}}\,\,.
$$
In the limit $r\rightarrow\infty$, we obtain $T=0$. So, this solution is asymptotically torsionless in comparison to the asymptotically flat solutions in GR.
For better presentation of our results, it is convenient to write the spatial part of  
(\ref{g}) in the following form
\begin{eqnarray}
\gamma_{ij}dx^idx^j=-\frac{5\beta e^{(\frac{c_1}{c_2-c_1})\tilde{r}^2}}{c_2+\frac{3}{2}e^{-2c_2}\tilde{r}^2}(d\tilde{r}^2+dz^2)+\tilde{r}^2d\tilde{\varphi}^2\label{gamma2},
\end{eqnarray}
through the following redefinitions of the  coordinates
\begin{eqnarray}
\tilde{z}=z\,\,,\\
\tilde{t}=e^{c1}t\,\,,\\
\tilde{\varphi}=c_0e^{-c_1}\varphi\,\,,\\
\tilde{r}=e^{c_2-c_1}r\,\,.
\end{eqnarray}
So, the general metric for static vacuum solution for viable model of $f(T)=T+\beta T^2$ is written in the form
\begin{eqnarray}
g_{\mu\nu}dx^\mu dx^\nu=d\tilde{t}^2-\gamma_{ij}dx^idx^j.
\end{eqnarray}
with the spatial metric, given in (\ref{gamma2}).

\subsection{General solution with finite values of $u(r)$ on the axis $r=0$}
In Weyl coordinates, we introduce the following general solution for metric function $u(r)$
\begin{eqnarray}
u(r)=\frac{1}{\pi}\int_{0}^{\pi}f(i r \cos \Theta)d\Theta\label{uj}
\end{eqnarray}
Further, we adopt the gauge $w(r)=r$, compatible with the usual vacuum solutions in GR. One good choice for arbitrary function $f$ is $f(\phi)=e^{\phi}$. Further, for obtaining  at least one non trivial solution (here TEGR case $f(T)=T$ is equivalent to the GR), we choose the viable model $f(T)=T+\beta T^2$. Now, we explicitly substitute the ansatz (\ref{uj}) in the field equations, in vacuum case. By integrating  this system, one obtains the following metric functions
\begin{eqnarray}
k_1(r)=c_1,\ \ k_2(r)=\frac{1}{2}\log(-\frac{5\beta}{3}\frac{1}{c_1+\int re^{-2J_0(r)}dr})\,\,.
\end{eqnarray}
Our motivation for introducing such explicit solution is that this solution is regular on axis $r=0$ and describes the general closed-form solution that corresponds to the model $f(T)=T+\beta T^2$. The solutions $k_1(r),k_2(r)$ can be unified in one form by the redefinition of the arbitrary constant $c_1$. Indeed, we can combine two solutions and write the following exact solution
\begin{eqnarray}
k(r)=c_2-\frac{1}{2}\log (c_1+\int re^{-2J_0(r)}dr)\,\,.
\end{eqnarray}
It is very interesting for us to observe that this solution is a constant torsion solution, due to (\ref{t}), in which it reads as  $T=e^{-2c_1}$. More precisely, it can be interpreted as an effective cosmological constant in the form of $\Lambda_{eff}=\frac{e^{-2c_1}}{2}$ in action. From positivity of the cosmological constant term, we obtain $0<c_1<\infty$.

\subsection{Solution with cosmological constant $\Lambda$}
In GR, there is a general family of cylindrically symmetric solutions for Einstein-Hilbert action with cosmological constant, introduced independently by Linet \cite{linet} and Tian \cite{tian},  abbreviated  as LT. With scalar field, there is another extension of the LT family introduced in \cite{ijmpa}. In this section we propose to find the solutions in the presence of the cosmological constant $\Lambda$. Again we examine $f(T)=T+\beta T^2$ but with a cosmological constant $\Lambda$, i.e
$$
f(T)=T+\beta T^2+\Lambda\,\,.
$$
The source of such solution can be a rod, thus it is appropriate to put the $\log$ Newtonian potential for $u(r)=\frac{2}{3}\log (\cos(\lambda r))$. Further, we adopt the gauge $w(r)=\lambda ^{-1} \sin(\lambda r)\cos^{1/3}(\lambda r),\lambda=\frac{1}{2}\sqrt{3\Lambda}$ \cite{tian}. In this case, the remaining component of the metric $k(r)$  for $\beta\leq \frac{9}{20\Lambda}$, and for small $\beta$ reads
$$
k(r)=\frac{1}{2}\log(\frac{-2\beta}{c_1+\int \frac{r}{\cos^{4/3}(\lambda r)}dr}),\ \ |\beta|<<1 \,\,.
$$
The torsion scalar here is variable and  reads
\begin{eqnarray}
T&=&-\frac{2\lambda}{3}\frac{\Sigma}{\left( \sin \left( \lambda\,r \right)  \right)}\,\,,\\
\Sigma &=& \frac{2}{3}\,\lambda\,\sin \left( \lambda\,r \right)  \left( 
\sqrt [3]{\cos \left( \lambda\,r \right) }- \left( \cos \left( \lambda
\,r \right)  \right) ^{7/3} \right) \int \!{\frac {r}{ \left( \cos
 \left( \lambda\,r \right)  \right) ^{4/3}}}{dr}+r \left( \cos \left( 
\lambda\,r \right)  \right) ^{2}-\frac{1}{4}\,r\\ \nonumber &+&\frac{2{\it c_1}}{3}\,\lambda\,\sqrt [3]{\cos
 \left( \lambda\,r \right) }\,\sin \left( \lambda\,r
 \right)-\frac{2{\it c_1}}{3}\,\lambda\, \left( \cos \left( \lambda\,r \right) 
 \right) ^{7/3}\,\sin \left( \lambda\,r \right)  \,\,.
\end{eqnarray}
 So, the metric in this case can be interpreted as the torsion based model of the common LT family in GR. Finally we mention here that the solutions obtained here have a dual family which can be obtained by imposing the following transformations
\begin{eqnarray}
t\rightarrow iz,\ \ z\rightarrow it,\ \ 2u\rightarrow u.
\end{eqnarray}
The first transformations are just the weak rotation of the coordinate $t$ or $z$. But the last one can be interpreted as the scaling invariance transformation of the potential function $u$. So, by applying these transformations we can obtain new families. Also by applying the Ehler's transformation with a ``NUT" parameter, we can obtain the dual stationary family of the solutions which can be interpreted as the NUTY family \cite{prd2005}.

\section{Conclusion}
Cylindrical symmetry in relativity turns out to be similar to spherical symmetry in many ways but quite different in others. Since modified theories of gravity are been used as alternative to GR, this aspect has also been extended. In this paper, $f(T)$ theory of gravity has been considered and some static solutions with cylindrical symmetry have been investigated. The general field equations is presented and a tetrads basis for cylindrical metric is assumed leading to four independent equations, linked with the energy density $\rho$, the radial pressure $p_r$ and the two latter, in the directions $\varphi$ and $z$, respectively, $p_\varphi$ and $p_z$. We specially focused our  attention on vacuum solutions, where we assumed $\rho=p_r=p_\varphi=p_z=0$.\par
We first assumed the algebraic function $f(T)=T$, the TEGR case. Expressions depending on the radial coordinate $r$ have been obtained for $k$, $u$ and $\omega$, leading to the line element (\ref{line1}). Assuming the suitable parametrizations, (25)-(30), we obtained the line element (31) which is formally similar to the Levi-Civita's static cylindrical symmetry solution in GR without $\pm$, but with the constant $m$ taking both positive and negative values. Still in this case, we shown that for $m=0$, the spacetime is characterized by the exterior metric of a cosmic string with the line element (32).\par
The second expression assumed is $f(T)=T+\beta T^2$, where the  parameters $k$, $u$ and $\omega$ are obtained in (34) yielding the line element (35).\par
At third step, we assumed a general solution with finite values of $u(r)$ on the axis $r=0$ and with the gauge $w(r)=r$, we obtained the coming expression of $k$. The interesting aspect here is that the solution is regular on the axis $r=0$ and describes the general closed-form solution for $f(T)=T+\beta T^2$. This is LT-like solution for $f(T)$ theory with cylindrical symmetry. \par
We finally took into account the cosmological constant, assuming $f(T)=T+\beta T+\Lambda$. In this case, for commodity, we used the $log$ Newtonian potential taking $u(r)=\frac{2}{3}log(\cos{(\lambda r)})$, and adopting the gauge $\omega(r)=\lambda^{-1}\sin{(\lambda r)}\cos^{1/3}{(\lambda r)}$, we obtained the exact expression for $k(r)$ for small $\beta$, and the torsion scalar as radial coordinate dependent.\par 
It appears from all this that as possible for spherical symmetry, static  cylindrical solutions can be obtained in Weyl coordinates in the framework of $f(T)$ theory. Note that the solutions are obtained here in the vacuum. As perspective, in a future work, we hope to take into account the matter content of the universe as, both isotropic and anisotropic fluid.

\vspace{0.5cm}
{\bf Acknowledgement:}  We thank a lot the referee for his useful suggestions. M. J. S. Houndjo  thanks CNPq/FAPES for financial support.

\end{document}